\documentclass{article}

\usepackage{PRIMEarxiv}

\usepackage[utf8]{inputenc} 
\usepackage[T1]{fontenc}    
\usepackage{hyperref}       
\usepackage{url}            
\usepackage{booktabs}       
\usepackage{amsfonts}       
\usepackage{nicefrac}       
\usepackage{microtype}      
\usepackage{lipsum}
\usepackage{fancyhdr}       
\usepackage{graphicx}       
\graphicspath{{media/}}     

\usepackage{xcolor}
\definecolor{maroon}{cmyk}{0, 0.87, 0.68, 0.32}
\definecolor{halfgray}{gray}{0.55}
\definecolor{ipython_frame}{RGB}{207, 207, 207}
\definecolor{ipython_bg}{RGB}{247, 247, 247}
\definecolor{ipython_red}{RGB}{186, 33, 33}
\definecolor{ipython_green}{RGB}{0, 128, 0}
\definecolor{ipython_cyan}{RGB}{64, 128, 128}
\definecolor{ipython_purple}{RGB}{170, 34, 255}

\usepackage{listings}
\lstdefinelanguage{iPython}{
    morekeywords={access,and,break,class,continue,def,del,elif,else,except,exec,finally,for,from,global,if,import,in,is,lambda,not,or,pass,print,raise,return,try,while},%
    morekeywords=[2]{abs,all,any,basestring,bin,bool,bytearray,callable,chr,classmethod,cmp,compile,complex,delattr,dict,dir,divmod,enumerate,eval,execfile,file,filter,float,format,frozenset,getattr,globals,hasattr,hash,help,hex,id,input,int,isinstance,issubclass,iter,len,list,locals,long,map,max,memoryview,min,next,object,oct,open,ord,pow,property,range,raw_input,reduce,reload,repr,reversed,round,set,setattr,slice,sorted,staticmethod,str,sum,super,tuple,type,unichr,unicode,vars,xrange,zip,apply,buffer,coerce,intern},%
    sensitive=true,%
    morecomment=[l]\#,%
    morestring=[b]',%
    morestring=[b]",%
    morestring=[s]{'''}{'''},
    morestring=[s]{"""}{"""},
    morestring=[s]{r'}{'},
    morestring=[s]{r"}{"},%
    morestring=[s]{r'''}{'''},%
    morestring=[s]{r"""}{"""},%
    morestring=[s]{u'}{'},
    morestring=[s]{u"}{"},%
    morestring=[s]{u'''}{'''},%
    morestring=[s]{u"""}{"""},%
    literate=
    {á}{{\'a}}1 {é}{{\'e}}1 {í}{{\'i}}1 {ó}{{\'o}}1 {ú}{{\'u}}1
    {Á}{{\'A}}1 {É}{{\'E}}1 {Í}{{\'I}}1 {Ó}{{\'O}}1 {Ú}{{\'U}}1
    {à}{{\`a}}1 {è}{{\`e}}1 {ì}{{\`i}}1 {ò}{{\`o}}1 {ù}{{\`u}}1
    {À}{{\`A}}1 {È}{{\'E}}1 {Ì}{{\`I}}1 {Ò}{{\`O}}1 {Ù}{{\`U}}1
    {ä}{{\"a}}1 {ë}{{\"e}}1 {ï}{{\"i}}1 {ö}{{\"o}}1 {ü}{{\"u}}1
    {Ä}{{\"A}}1 {Ë}{{\"E}}1 {Ï}{{\"I}}1 {Ö}{{\"O}}1 {Ü}{{\"U}}1
    {â}{{\^a}}1 {ê}{{\^e}}1 {î}{{\^i}}1 {ô}{{\^o}}1 {û}{{\^u}}1
    {Â}{{\^A}}1 {Ê}{{\^E}}1 {Î}{{\^I}}1 {Ô}{{\^O}}1 {Û}{{\^U}}1
    {œ}{{\oe}}1 {Œ}{{\OE}}1 {æ}{{\ae}}1 {Æ}{{\AE}}1 {ß}{{\ss}}1
    {ç}{{\c c}}1 {Ç}{{\c C}}1 {ø}{{\o}}1 {å}{{\r a}}1 {Å}{{\r A}}1
    {€}{{\EUR}}1 {£}{{\pounds}}1
    {^}{{{\color{ipython_purple}\^{}}}}1
    {=}{{{\color{ipython_purple}=}}}1
    {+}{{{\color{ipython_purple}+}}}1
    {*}{{{\color{ipython_purple}$^\ast$}}}1
    {/}{{{\color{ipython_purple}/}}}1
    {+=}{{{+=}}}1
    {-=}{{{-=}}}1
    {*=}{{{$^\ast$=}}}1
    {/=}{{{/=}}}1,
    literate=
    *{-}{{{\color{ipython_purple}-}}}1
     {?}{{{\color{ipython_purple}?}}}1,
    identifierstyle=\color{black}\ttfamily,
    commentstyle=\color{ipython_cyan}\ttfamily,
    stringstyle=\color{ipython_red}\ttfamily,
    keepspaces=true,
    showspaces=false,
    showstringspaces=false,
    rulecolor=\color{ipython_frame},
    frame=single,
    frameround={t}{t}{t}{t},
    framexleftmargin=6mm,
    numbers=left,
    numberstyle=\tiny\color{halfgray},
    backgroundcolor=\color{ipython_bg},
    basicstyle=\scriptsize,
    keywordstyle=\color{ipython_green}\ttfamily,
}

\title{GMP-Featurizer: A parallelized Python package for efficiently computing the Gaussian Multipole features of atomic systems}

\author{
  Xiangyun Lei \\
  Toyota Research Institute \\
  Los Altos, CA\\
  \texttt{ray.lei@tri.global} \\
   \And
  Joseph Montoya \\
  Toyota Research Institute  \\
  Los Altos, CA\\
  \texttt{joseph.montoya@tri.global} \\
}

\begin{document}
\maketitle



\section{Summary}
\label{sec:introduction}
GMP-Featurizer is a lightweight, accurate, efficient, and scalable software package for calculating the Gaussian Multipole (GMP) features \cite{GMP} for a variety of atomic systems with elements across the periodic table. Starting from the GMP feature computation module from AmpTorch \cite{amptorch}, the capability of GMP-Featurizer has since been greatly improved, including its accuracy and efficiency, as well as the ability to parallelize on different cores, even machines. Moreover, this python package only has very few dependencies that are all standard python libraries, plus cffi for C++ code interfacing and Ray \cite{Ray} for parallelization,  making it lightweight and robust. A set of unit tests are designed to ensure the reliability of its outputs. A set of extensive examples and tutorials, as well as two sets of pseudopotential files (needed for specifying the GMP feature set), are also included in this package for its users. Overall, this package is designed to serve as a standard implementation for chemical and material scientists who are interested in developing models based on GMP features. The source code for this package is freely available to the public under the Apache 2.0 license.

\section{Statement of Needs}
\label{sec:SON}
Representing the local and global environments in atomic systems in a descriptive and efficient way has been an important research top in the chemistry, chemical engineering, and material science communities. Having good representations, or features, of chemical environments has proven to be vital for building reliable machine learning (ML) models. These models can accurately predict properties of atomic systems, and in limited cases have even been used for discovering or designing new chemicals and materials\cite{zuo_accelerating_2021,collins_accelerated_2017}. So far, scientists and researchers have designed featurization schemes like the atom-centered symmetry function (ACSF) \cite{BehlerParrinello}, the smooth overlap of atomic positions (SOAP) \cite{SOAP} and the Gaussian Momentum \cite{Gaussian_Momentum} schemes. More recently, graph representation and ML models based on them (e.g. MEGNet\cite{MEGNet}, CGCNN\cite{CGCNN}) have been successful. Gaussian Multipole, or GMP \cite{GMP}, is a recently developed scheme of featurizing local chemical environments, i.e. the chemical characteristics of spaces near individual atoms in molecules and crystal structures. GMP approximates underlying local electronic environments (e.g. approximated distribution of local electron cloud) using multipole expansion, the theory of which is explicated in a prior publication\cite{GMP}. The featurization scheme is flexible, depending only on prior assumptions of atomic identity and position, and it is therefore applicable to various atomic systems (molecules, nano-particles, periodic crystals, etc.) in which atomic arrangements are known. Feature computation is fast, and the representation accuracy is systematically improvable. Moreover, thanks to the deep connection between the GMP features and physics, we have previously shown that ML models based on these features are transferable\cite{GMP}. With these characteristics, GMP featurization could be useful to a broad audience for future research in chemistry and materials science. Therefore, having lightweight, reliable, and open software that can calculate these features in a fast and accurate way is desirable.

\section{Overview}
\label{sec:Overview}
The GMP-Featurizer package is mainly written in C++ and python. C++ is used for the underlying computation module for speed, and python used for an intuitive and readable API for scientists and researchers in the community to use. Although this package does not explicitly depend on ASE \cite{ase-paper} and pymatgen \cite{pymatgen}, two of the most widely used python libraries in the chemistry and materials science communities, APIs are provided so atomic systems defined in these libraries can be easily read by gmp-featurizer. On top of that, Ray is used to parallelize the feature computation, and the parallelization efficiency is close to 100\%.  Overall, this package is designed to be lightweight, easy to use, fast and accurate.

The main inputs of the workflow are a python dictionary that contains the necessary hyperparameters for defining the desired GMP feature set, and a list of atomic systems that needs to be featured. The native way of defining atomic systems is simply a python dictionary that contains information like `lattice vectors`, `atom positions`, `atom types`, etc. As mentioned, the package also supports both ASE Atoms and pymatgen Structure objects with pre-defined converters. This capability is extensible to other formats with custom-made converters. It also supports the featurization of disordered atomic structures, which is unsupported by many popular featurization methods. Please refer to section \ref{sec:Examples} for more details. The output is simply a list of dictionaries containing the resulting features, and their derivatives if requested, for the atomic structures.

By default, the package computes GMP features at each atom position, but it can also be used to compute the features at any set of reference points inside the atomic system by providing a list of the positions of interest for each atomic structure. Users can also specify the number of cores for parallel computing. Moreover, computed results can be cached locally for convenient reprocessing of datasets, e.g. after augmentation or modification. Two sets of standard pseudopotential files are also provided, which are necessary to specify GMP feature sets, but may be difficult to collect from either commercial or open-source density functional theory systems. Lastly, a series of tutorials are provided in the repository to help users with quick starting and understanding the various features of the codebase.

\section{Examples}
\label{sec:Examples}
Here we provide a quick example to showcase the interface of this package. For more detailed tutorials, please refer to the examples included in the repository.

To started, one must import the featurizer module, which is the main interface to the GMP feature computation routine. The list of atomic systems also must be prepared along with the appropriate converter, if needed.
\begin{lstlisting}[language=iPython]
from GMPFeaturizer import GMPFeaturizer, ASEAtomsConverter, PymatgenStructureConverter
import pickle

# Say the example data is a list of ase atoms object
with open("./example.p", "rb") as f:
    systems = pickle.load(f)

# initialize the converter, in this case it's the converter for ASE atoms objects
# There is also a pre-existing converter for pymatgen Structure objects as well
converter = ASEAtomsConverter()
# converter = PymatgenStructureConverter()
\end{lstlisting}

Then, a dictionary of hyperparameters for specifying the GMP feature set must be defined, with which the featurizer can be initialized. 

\begin{lstlisting}[language=iPython]
GMPs = {
    "GMPs": {   
        "orders": [-1, 0, 1, 2], 
        "sigmas": [0.1, 0.2, 0.3]   
    },
    # path to the pseudopotential file
    "psp_path": "<path>/NC-SR.gpsp", 
    # basically the desired accuracy of the features
    "overlap_threshold": 1e-16, 
}
featurizer = GMPFeaturizer(GMPs=GMPs, calc_derivatives=True)
\end{lstlisting}

Here, the list of orders corresponds to the orders of the MCSH angular probe functions, and the list of sigmas corresponds to the sigmas of the Gaussian radial probes (see Ref\cite{GMP} for details). The resulting feature set is defined by the Cartesian product of the lists of orders and sigmas. The exception is order -1, which corresponds to local electron density, so the sigma value does not matter. Hence, there is only one feature for order -1.

Therefore, with the above setting, the feature set looks like 

$[(-1, 0), (0, 0.1), (0, 0.2), (0, 0.3), (1, 0.1), (1, 0.2), (1, 0.3), (2, 0.1), (2, 0.2), (2, 0.3)]$

where again the first number is the order of the MCSH angular probe, and the second number is the sigma of the Gaussian radial probe. Note that by setting \texttt{calc\_derivatives=True}, we are asking the featurizer to compute the GMP feature derivative w.r.t. the atom positions as well. By default, the user won't need to specify the local cutoff distance. The distance will be inferred by the \texttt{overlap\_threshold} setting.

At this point, one can compute and access the desired set of GMP features for the atomic systems simply by

\begin{lstlisting}[language=iPython]
result = featurizer.prepare_features(systems, cores=5, converter=converter)

gmp_features = [entry["features"] for entry in result]
gmp_feature_derivatives = [entry["feature_primes"] for entry in result]
\end{lstlisting}

Here, \texttt{cores} specifies the number of cores for parallel computing. 

Therefore, the entire script for computing the GMP features of a given list of atomic systems is shown below:

\begin{lstlisting}[language=iPython]
from GMPFeaturizer import GMPFeaturizer, ASEAtomsConverter, PymatgenStructureConverter
import pickle

# Say the example data is a list of ase atoms object
with open("./example.p", "rb") as f:
    systems = pickle.load(f)
converter = ASEAtomsConverter()

GMPs = {
    "GMPs": {   
        "orders": [-1, 0, 1, 2], 
        "sigmas": [0.1, 0.2, 0.3]   
    },
    "psp_path": "<path>/NC-SR.gpsp", 
    "overlap_threshold": 1e-16, 
}
featurizer = GMPFeaturizer(GMPs=GMPs, calc_derivatives=True)
result = featurizer.prepare_features(systems, cores=5, converter=converter)

gmp_features = [entry["features"] for entry in result]
gmp_feature_derivatives = [entry["feature_primes"] for entry in result]
\end{lstlisting}


\section*{Acknowledgments}
This work was supported by the Energy and Materials Division of the Toyota Research Institute. The authors acknowledge Jens Hummelshøj for helpful discussions regarding unit-testing frameworks.

\bibliographystyle{unsrt}  
\bibliography{references}

\end{document}